\documentclass[dvips]{article}
\usepackage{icrctc07}
\title{Discovery of Very High Energy Gamma-Rays from the Distant Flat Spectrum Radio Quasar 3C~279 with the MAGIC Telescope }
\shorttitle{Discovery of VHE Gamma-Rays from 3C~279}
\authors{Masahiro Teshima,$^{1}$ Elisa Prandini,$^{2}$ Rudolf Bock,$^{1,2}$ Manel Errando,$^{3}$ Daniel Kranich,$^{4}$
Pratik Majumdar,$^{1}$ Daniel Mazin,$^{3}$ Elina Lindfors,$^{5}$
Eckart Lorenz,$^{4,1}$ Mos\`{e} Mariotti,$^{2}$ Villi
Scalzotto,$^{2}$ and Robert Wagner$^{1}$ for the MAGIC
Collaboration}
\shortauthors{Teshima et al.}
\afiliations{$^1$ Max-Planck-Institut f\"ur Physik, D-80805 M\"unchen, Germany\\
$^2$ Universit\`a di Padova and INFN, I-35131 Padova, Italy\\
$^3$ Institut de F\'{\i}sica d'Altes Energies, E-08193 Barcelona, Spain\\
$^4$ ETH Zurich, CH-8093 Zurich, Switzerland\\
$^5$ Tuorla Observatory, Turku University, FI-21500 Piikki\"o,
Finland}
\email{elisa.prandini@pd.infn.it \& robert.wagner@mppmu.mpg.de}

\abstract{The quasar 3C~279 is one of the best-studied flat
spectrum radio quasars.  It is located at a comparatively large
redshift of $z=0.536$: $E>100$~GeV observations of such distant
sources were until recently impossible both due to the expected
steep energy spectrum and the expected attenuation of the
$\gamma$-rays by the extragalactic background light. Here we
present results on the observation of 3C~279 with the MAGIC
telescope in early 2006. We report the detection of a significant
very high energy $\gamma$-ray signal in the MAGIC energy range on
the observation night of 2006 February 23.}

\begin{document}
\maketitle
\section{Introduction}
3C~279 (RA=12$^\mathrm{h}$56$^\mathrm{h}$11.1$^\mathrm{s}$,
Dec=$-$5$^\circ$47$^\mathrm{m}$22$^\mathrm{s}$) was the first blazar
discovered in $\gamma$-rays with the {\it Compton Gamma-Ray Observatory}
\cite{Egret1991}. It is an exceptionally bright and variable
source at various energy bands. In the optical, typical variations
of 2$^\mathrm{m}$ have been observed, while during flares, up to
8$^\mathrm{m}$ were reached \cite{Webb-90}. Variations
were observed on day-scale \cite{Hartman-01b}, are mostly coherent at
all frequencies, and the spectral energy distribution (SED)
hardens when the source brightens \cite{Maraschi-94}. Typical flux
variations by factors of $\sim 20$ in the GeV and $\sim 5 -
10$ in the IR-to-UV have been found \cite{Klaas}.
The strongest variability occurs on timescales of a few weeks to
$\sim 6$ months.

Blazars are thought to be supermassive black holes in the centers
of galaxies accreting matter. They possess strongly collimated,
ultra-relativistic plasma outflows (jets), aligned closely to the
observer's line of sight. Thus, their SEDs are almost entirely
dominated by the jet emission. In $\nu F_\nu$ representations,
they are characterized by two distinct nonthermal components. The
lower energy bump is commonly ascribed to synchrotron radiation
emitted by relativistic electrons. There is less agreement about
the origin of the high-energy bump. In leptonic acceleration
models, inverse Compton (IC) scattering of synchrotron
(self-synchrotron Compton, SSC, models) or ambient photons
(external-inverse Compton, EIC, models) on high energy electrons
explain the MeV-to-TeV radiation. While BL Lac objects can often
successfully be described by SSC models, the more luminous Flat
Spectrum Radio Quasars (FSRQ) usually are modeled requiring
external components, particularly in the $E>100$~keV regime
\cite{hartman}, although the existence of EIC components is
questioned e.g. by \cite{g,l5,l6}.

Since its discovery in $\gamma$-rays, 3C~279 was extensively
studied, in particular also during various multi-wavelength
campaigns
\cite{Hartman-01b,Maraschi-94,hartman,Wehrle-98,bach,Boett-07}.
The broadband SED extends from radio frequencies to the $\gamma$
regime and is, with a comparatively low synchrotron peak, fairly
typical of FSRQs. Correlations between the two peaks, which are
expected in leptonic models, were partially observed
\cite{hartman}, although no consistent patterns were found. The
SED of this source is rather complicated, and in spite of detailed
observations, still poorly understood. Observations of the
high-energy part of the spectrum are complicated by the fact that
terrestrial and satellite-borne instruments together can, at
present, not fully cover the frequency range of the high-energy
bump. Additionally, the very high energy (VHE, defined by
$E>100$~GeV) spectrum is in part suppressed from interactions with
the extragalactic background light (EBL). This modification is, of
course, strongly dependent on the source distance. Thus, distant
VHE $\gamma$-ray sources represent an excellent tool for
determining the 0.3 to 30 $\mu$m EBL (e.g., \cite{EBL2}), which at
redshift $z=0$ was observed by various satellite experiments,
although direct measurements suffer from huge foreground
contaminations by light contributions from the solar system and
our galaxy.  As of now, several different EBL models have been
proposed \cite{EBL2,EBL,EBL3}. A precise measurement of the energy
spectrum of 3C~279 is crucial for two reasons: With a
state-of-the-art EBL model, emission models for 3C~279 can be
tested in detail. By using conservative arguments on particle
acceleration mechanisms, the $\gamma$-ray emission of 3C~279 also
permits to formulate stringent constraints on the EBL level.

All extragalactic VHE $\gamma$-ray sources detected so far are of
the BL~Lac type. These objects are the low-luminosity counterparts
of the FSRQ class of AGNs \cite{Urry-Padovani}, with their
synchrotron peaks shifted to higher energies \cite{Fossati}.
BL~Lac objects have been detected aplenty recently in the VHE
range: 17 BL~Lacs have been found so far,\footnote{See
http://www.mppmu.mpg.de/$\sim$rwagner/sources/ for an up-to-date
list} reaching to redshifts of $z=0.212$ \cite{1011}, plus M87,
assumed to be a misaligned BL~Lac. The sample includes also
blazars as PG\,1553+113 \cite{153} with its extremely soft energy
spectrum and recently discovered BL~Lacertae itself
\cite{bllacertae}, which is the first ``low-peaked BL~Lac object''
($\nu_\mathrm{synchr}<10^{14}$~Hz). Both observations were largely
made possible from the exceptionally low energy threshold of the
MAGIC telescope.

In January 2006, 3C~279 was found in a high optical state,
brightening to $14^\mathrm{m}5$ in the
$R$-band.\footnote{Long-term optical monitoring data are available
at http://users.utu.fi/kani/1m/} A WEBT campaign in early 2006
\cite{Boett-07,boettcher} and multi-wavelength campaigns during
both periods \cite{bach} were performed in order to get further
information on the temporal and spectral properties of 3C~279.

Up to recently, a VHE $\gamma$-ray detection of 3C~279 was
prevented by its high redshift of $z=0.536$. The resulting cutoff
due to EBL attenuation is expected at around $E\approx200$~GeV.
The MAGIC (Major Atmospheric Gamma-ray Imaging Cerenkov) telescope
\cite{MAGIC1} is currently the largest single-dish (17~m diameter)
Imaging Atmospheric Cerenkov Telescope, located on the Canary
Island of La Palma. With its low trigger threshold ($50-60$~GeV at
low zenith angles) is best suited for observing the lower part of
the VHE range of distant AGN spectra. At the geographical latitude
of MAGIC ($28^\circ45'$~N), 3C~279 can be observed under medium
zenith angles (above 34$^\circ$), with an accordingly increased
observation threshold.

\section{Observations and Analysis}
3C~279 was observed from late January to April 2006 and in January
2007. Simultaneous optical $R$-band observations were carried out
using the 1.03~m telescope at the Tuorla Observatory, Finland, and
the 35~cm KVA telescope on La Palma. Here we report results from
the 2006 observations.

The VHE $\gamma$-ray observations were performed in the ON-OFF
mode. ON data were collected while pointing directly to the
source, while OFF data, necessary for the background estimation,
were recorded by pointing to a nearby region of the sky. The OFF
region, in which no $\gamma$-ray source is expected,  was chosen
to have a comparable zenith angle distribution and night sky
conditions. From 2006 January 29 to 2006 April 4, 3C~279 was
observed for 14.9~hours. In addition, 3.9 hours of OFF data were
recorded. Data taken during non-optimal weather conditions or
affected by hardware problems were excluded from the analysis.
This concerned 5.2 hours worth of ON data. The remaining events
were calibrated \cite{gaug} and analyzed using the standard MAGIC
analysis chain \cite{Magic-Software,wagner05}. Briefly, the
analysis proceeds by reducing the image to a single light cluster
by removing noise, calculating image parameters
\cite{Hillas_parameters}, and using a multi-variant method to
discriminate $\gamma$-like events from the dominant hadronic
background. For the latter step, we use the Random Forest (RF)
method \cite{bock}, which combines the image parameters into a
single test statistic, called hadronness, based on training
samples of both Monte Carlo generated $\gamma$
\cite{knapp04,majumdar05} and real background events. The excess
events were identified using the classical ALPHA approach by
subtracting suitably normalized OFF data from the ON events. The
significance of any excess was calculated according to eq.~17 in
\cite{li83}. The energy estimation for each event was performed
using a RF method, too, leading to a reasonably constant energy
resolution of $\approx$23\% above 150~GeV.

\vspace*{-.1cm}
\section{Results}

\begin{figure}
\begin{center}
\includegraphics[width=\linewidth]{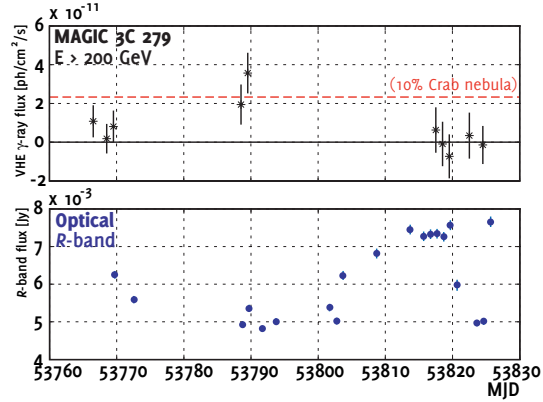}
\caption{MAGIC $E>200$~GeV $\gamma$-ray (top) and optical $R$-band
(bottom) light curves obtained for 3C~279 in early 2006.}
\label{lightcomb}
\vspace*{-.2cm}
\end{center}
\end{figure}

The data were subjected to three independent analysis chains,
which obtain compatible results w.r.t. each other. A standard
analysis with an energy threshold of $\approx 200$~GeV was used to
infer the $E>200$~GeV light curve given in Figure \ref{lightcomb}.
Also shown is a $R$-band optical light curve for ten MAGIC
observation nights from 2006 January 31 to 2006 March 31. The
typical observation time per night is around one hour. While in
most of the nights $\gamma$-ray fluxes compatible with zero were
observed, during the 2006 February 22 observations a marginal
signal was seen. In the night of 2006 February 23 we found a clear
$\gamma$-ray signal with an integrated photon flux
$F(E>200\,\mathrm{GeV})=(3.5\pm0.8)\times10^{-11}\,\mathrm{cm}^{-2}\,\mathrm{s}^{-1}$.
The source was observed for 62 minutes (MJD
53789.1633$-$53789.2064) at zenith angles between $35^\circ$ and
$38^\circ$. The stable event rate during this observation allows
classifying it as ``dark night'' rate, although minimal moonshine
was present.

\begin{figure}
\begin{center}
\includegraphics[width=\linewidth]{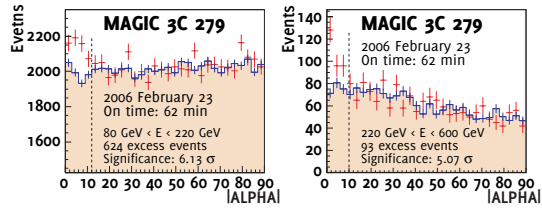}
\caption{The $\gamma$-ray signal below (left) and above (right)
220 GeV on 2006 February 23. Showers which are aligned parallel to
the telescope axis have small ALPHA values. To remove residual
hadronic background, OFF data were normalized to the ON
distribution between $20^\circ<\mathrm{ALPHA}<80^\circ$ and
subtracted from it. The structure visible in the left ALPHA plot
in the OFF data is statistically compatible with a flat
distribution. } \label{alpha}
\end{center}
\end{figure}

A low-energy analysis, exploiting the timing properties of the air
shower images with a lower analysis threshold, $\approx 110$~GeV,
was used to calculate the significance of the found excess.
Threshold is defined as the maximum of the energy distribution of
the accepted events, {\em viz.} showers with energies down to
$\approx 80$~GeV are included in the sample. Both analyses show
consistent results from $E\geq 200$~GeV on.
After subjecting the ON data to an ALPHA cut inferred from
training on an (independent) Crab nebula data sample recorded at
similar zenith angles, and subtracting a properly normalized OFF
distribution in the signal region, 624 and 93 excess events remain
between 80 and 220 GeV and between 220 and 600 GeV, respectively
(Fig. \ref{alpha}). The data were separated into these two
independent samples because of the very different $\gamma$/hadron
separation powers in the two energy regions: Adding the
highly-enriched, low statistics high-energy $\gamma$ sample to the
large statistics low-energy sample would spoil the overall
significance calculation. The resulting significances are
$6.1\sigma$ in the low energy region and $5.1\sigma$ in the high
energy region; the excess is compatible with a point-like source
and its position is consistent, within statistical uncertainties,
with the 3C~279 position. The signal at low energies exceeds the
intensity of the Crab nebula, while for $E>200$~GeV it is
$\approx$15\% of the Crab nebula flux. The spectral analysis of
the data is ongoing. Our $\gamma$-ray detection was not
accompanied by an optical flare or by particularly high flux
levels or outbursts in X-rays \cite{XTE,Krimm}.

\section{Conclusions}
In  observations triggered by a high optical state of the flat
spectrum radio quasar 3C~279, MAGIC was able to detect a highly
significant VHE $\gamma$-ray signal from this source, well-known
and much studied at lower energies. This discovery is a crucial
step forward for VHE $\gamma$-ray astronomy in various contexts:
\begin{itemize}
\vspace*{-.2cm} \item The distance over which astrophysical
objects can be observed at VHE $\gamma$-ray energies was enlarged
substantially---when the photons recorded during our observations
left 3C~279, the age of the Universe was only 8.4 Gyr.
\vspace*{-.2cm} \item For the first time, VHE $\gamma$-rays from a
flat spectrum radio quasar were detected. These objects are very
luminous, but are not expected to show intense $E>100$~GeV
emission.
With 3C~279 and BL~Lacertae, objects of all classes comprising the
``blazar sequence'' \cite{Fossati} have now been detected in VHE
$\gamma$-rays.
\vspace*{-.2cm} \item A precise measurement of the energy spectrum
in the VHE region may allow for stringent constraints on the EBL
density.
\end{itemize}

\section*{Acknowledgments}
We thank the IAC for the excellent working conditions in La Palma.
The support of the German BMBF and MPG, the Italian INFN and
Spanish CICYT is gratefully acknowledged. This work was also
supported by ETH Research Grant TH 34/04 3 and by the Polish MNil
Grant 1P03D01028. Further we are grateful to the {\it RXTE}-ASM
and {\it Swift}/BAT teams for providing their X-ray monitoring
data and in particular to Hans Krimm for quickly analyzing the
2006 February 23 BAT data.

\end{document}